\documentclass[journal, twocolumn, lettersize]{IEEEtran}

\usepackage{graphicx} 
\usepackage{graphics}
\usepackage{multirow}
\usepackage{textcomp}
\usepackage{array}
\usepackage[labelformat=simple]{subcaption}

\usepackage{xcolor}
\usepackage{tabularx} 
\newcolumntype{Y}{>{\centering\arraybackslash}X} 
\usepackage{booktabs}

\usepackage{amsmath,amssymb,amsfonts}
\usepackage{bm}

\usepackage[algo2e, ruled, vlined, linesnumbered]{algorithm2e}

\usepackage{amsthm}
\theoremstyle{definition}

\SetKwProg{Fn}{Function}{}{end}

\SetKwInput{KwInit}{Init}
\SetAlgoSkip{}

\newcolumntype{M}[1]{>{\centering\arraybackslash}m{#1}}

\DeclareSymbolFont{symbolsC}{U}{txsyc}{m}{n}
\DeclareMathSymbol{\notniFromTxfonts}{\mathrel}{symbolsC}{61}

\usepackage{titlesec}

\titlespacing{\subsection}{0pt}{1.5ex plus .2ex minus .2ex}{0.5ex plus .2ex}

\title{Predictive Lightweight MARL for Resilient Coverage in Sparse-Signaling Aerial Networks}

\author{Chuan-Chi~Lai,~\IEEEmembership{Member,~IEEE}, and Ang-Hsun~Tsai,~\IEEEmembership{Member,~IEEE}
    \IEEEcompsocitemizethanks{        
        \IEEEcompsocthanksitem{
            This research was supported by the National Science and Technology Council, Taiwan, R.O.C., under Grant Nos. \mbox{NSTC 114-2221-E-194-062-,} NSTC 115-2221-E-194-042-MY2, and \mbox{NSTC 115-2221-E-035-050-.} This work was also partially supported by the Advanced Institute of Manufacturing with High-tech Innovations (AIM-HI) from the Featured Areas Research Center Program within the framework of the Higher Education Sprout Project by the Ministry of Education (MOE) in Taiwan. In addition, this work was sponsored by Feng Chia University under Grant 25H00812. \emph{(Corresponding author: Chuan-Chi~Lai.)}}
        \IEEEcompsocthanksitem{
            Chuan-Chi~Lai is with the Department of Communications Engineering, National Chung Cheng University, Minxiong Township, Chiayi County 621301, Taiwan, and also with the Advanced Institute of Manufacturing with High-tech Innovations (AIM-HI), National Chung Cheng University, Minxiong Township, Chiayi County 621301, Taiwan (e-mail: chuanclai@ccu.edu.tw).}
        \IEEEcompsocthanksitem{
            Ang-Hsun~Tsai is with the Department of Communications Engineering, Feng Chia University, Taichung 407102, Taiwan.}
        \IEEEcompsocthanksitem{\copyright~2026 IEEE.  Personal use of this material is permitted. For any other uses, permission must be obtained from IEEE.}
    }
}

\IEEEtitleabstractindextext{%
\begin{abstract}
    This letter proposes the Predictive Lightweight Multi-Agent Reinforcement Learning (PL-MARL) framework to ensure resilient coverage in bandwidth-constrained UAV swarms. To counter coordination collapse caused by sparse signaling and information aging, we introduce a Kinematic-Aware Inference Engine that proactively reconstructs neighbor trajectories via physical priors. This approach enables an efficient computation-for-communication trade-off, decoupling structural resilience from signaling frequency. Simulations confirm that PL-MARL maintains superior coverage and mission continuity under extreme signaling scarcity and node failure. Our results validate proactive inference as a scalable, low-latency solution for robust aerial coordination, effectively minimizing control overhead to preserve spectrum for payload services while ensuring resilience against interference.
\end{abstract}

\begin{IEEEkeywords}
Sparse signaling, Age of Information (AoI), Spatio-temporal inference, Resilient coverage, UAV swarms, Multi-agent reinforcement learning.
\end{IEEEkeywords}}
\begin{document}

\maketitle
\IEEEdisplaynontitleabstractindextext

%
\IEEEpeerreviewmaketitle

\section{Introduction}
\label{sec:intro}

\IEEEPARstart{I}{n} the 6G era, \textit{Unmanned Aerial Vehicle} (UAV) networks are pivotal for ubiquitous connectivity. Seamless coordination requires continuous state exchange, yet conventional \textit{Multi-Agent Reinforcement Learning} (MARL) schemes rely on high-frequency synchronization. In bandwidth-constrained environments, this overhead exhausts spectral resources and aggravates co-channel interference, causing severe information aging \cite{Pham2022, Yates2021}. While \textit{Digital Twin} (DT) frameworks \cite{Zhou2026DT} provide a stable training backplane, decentralized execution remains hindered by the difficulty of maintaining state awareness under sparse signaling.

Recent research explores swarm intelligence through graph-enhanced planning \cite{Du2024}, DT-enhanced resource allocation \cite{Luo2024}, and secure communications \cite{Tang2025}. While \textit{Quantum-Assisted Frameworks} \cite{Zhang2026Quantum} offer algorithmic potential, their specialized hardware dependencies are impractical for resource-constrained UAVs. 
More critically, state-of-the-art frameworks utilizing spatio-temporal attention for partial observability fundamentally rely on frequent signaling regimes; they lack the onboard generative capacity to synthesize unobserved states during prolonged silence, rendering them untenable in extreme \textit{Flying Ad-hoc Networks} (FANETs).
Specifically, our prior work, \textit{Topology-Aware Graph MAPPO} (TAG-MAPPO) \cite{lai2026resilienttopologyawarecoordinationdynamic}, achieves resilient reconfiguration via centralized graph aggregation but remains bound to continuous state synchronization for topological observability. 
Consequently, in signaling-sparse environments, these models face coordination collapse; lacking the capacity to synthesize unobserved states, they leave a critical gap between delayed feedback and the real-time requirements of agile aerial control. By contrast, this study shifts the paradigm from reactive synchronization to proactive, physics-guided state reconstruction, enabling robust coordination where prior methodologies fundamentally fail.

To bridge the aforementioned gap, this letter proposes the \textit{Predictive Lightweight Multi-Agent Reinforcement Learning} (PL-MARL) framework. Our core contribution is a Kinematic-Aware Inference Engine that functions as a local generative unit. It proactively reconstructs neighbor trajectories via physical motion priors to decouple decentralized policy execution from communication frequency. Unlike black-box recurrent baselines hindered by computational bottlenecks, this mechanism enables sub-millisecond inference and subverts sequential processing limitations. Furthermore, we introduce a topology-aware graph attention mechanism that balances computational efficiency with mission reliability. By deliberately trading redundant topological cohesion for an expanded spatial footprint, PL-MARL mitigates the coverage collapse observed in traditional reactive models and ensures graceful performance degradation. Finally, empirical results validate that PL-MARL establishes a robust control backbone capable of sustaining mission continuity under the dual pressure of prolonged sparse signaling and abrupt node failure. This confirms its deployment readiness for latency-aware aerial operations, even when physical feedback loops are effectively severed.

\section{System Model and Problem Formulation}
\begin{figure}[!t]
    \centering
    \includegraphics[width=.8\linewidth]{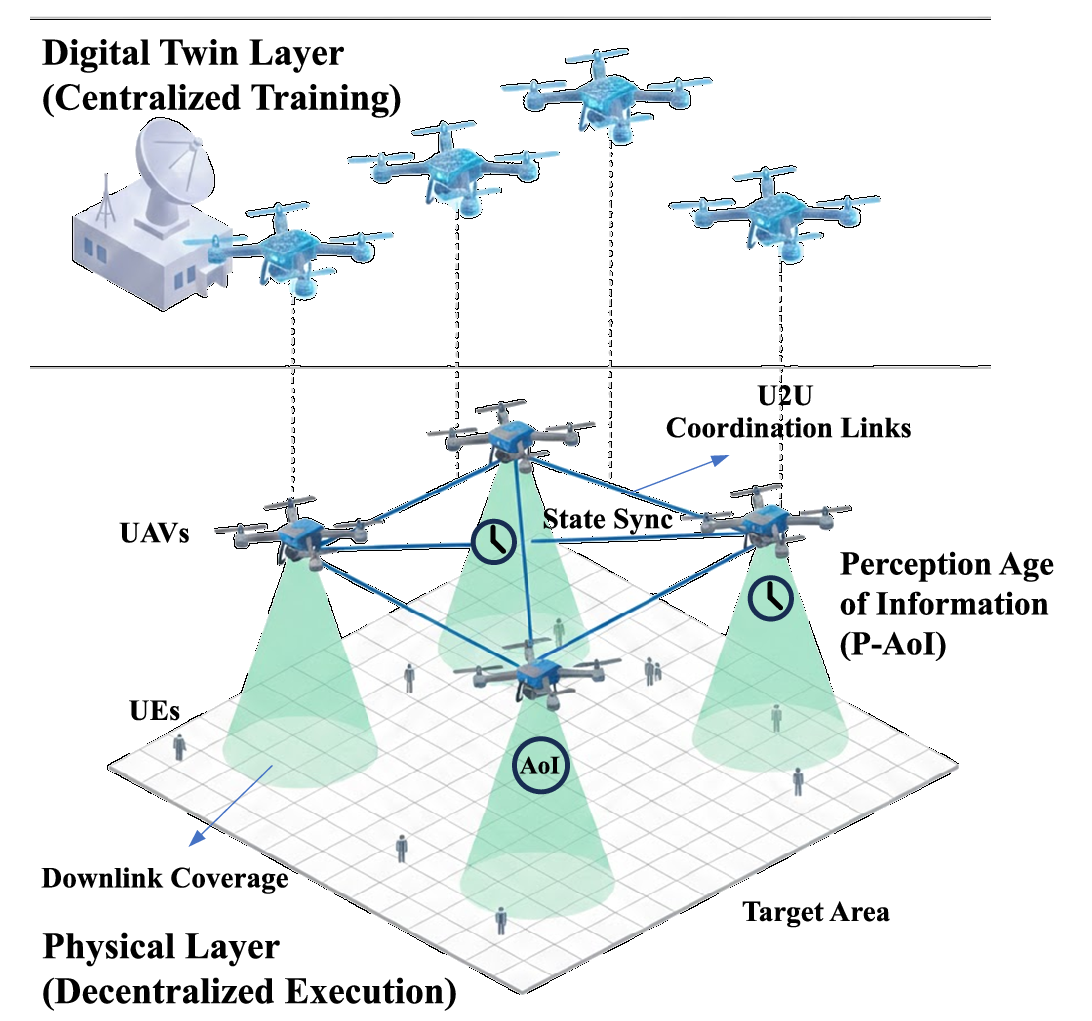}
    \caption{Architecture of the terminal-edge cooperative DT framework for signaling-efficient aerial networks.}
    \label{fig:system_scenario}
\end{figure}

We consider a UAV swarm $\mathcal{N} = \{1, \dots, N\}$ performing 3D coverage, where each node $i$ maintains state $\mathbf{s}_i(t) = [\mathbf{p}_i(t), \mathbf{v}_i(t)] \in \mathbb{R}^6$. As illustrated in Fig.~\ref{fig:system_scenario}, we implement a decoupled DT architecture. A ground server facilitates centralized training to mitigate non-stationarity. Once deployed, each UAV acts as an autonomous edge node to execute decentralized control. However, during sparse signaling, agents must rely on delayed historical information, leading to severe perception gaps within the communication range $R_{\mathrm{comm}}$.

UAVs communicate over a shared channel where a link requires the \textit{Signal-to-Interference-plus-Noise Ratio} (SINR) to exceed a threshold $\Gamma$. To mitigate interference in dense swarms, inter-UAV updates are restricted to a sparse interval $T_{\mathrm{up}} \in \mathbb{Z}^+$. Node $i$ broadcasts its true state $\mathbf{s}_i(t)$ only at $t = k T_{\mathrm{up}}$, where $k \in \mathbb{Z}_{\ge 0}$ is the index of the signaling cycle. Information staleness is quantified by the \textit{Perception Age of Information} (P-AoI):
\begin{equation}
\label{eq:p_aoi}
\Delta_{ij}(t) = \begin{cases}
        1, & t = k T_{\mathrm{up}} \\
        t - k T_{\mathrm{up}}, & t \in (k T_{\mathrm{up}}, (k+1) T_{\mathrm{up}})
    \end{cases}
\end{equation}
A larger $T_{\mathrm{up}}$ explicitly induces higher perception uncertainty.

We formulate the coordination as a \textit{Decentralized Partially Observable Markov Decision Process} (Dec-POMDP). At timestep $t$, node $i$ selects a continuous action $\mathbf{a}_i(t)$ (i.e., acceleration) using its available local perception $\mathcal{O}_i(t) = \{\mathbf{s}_i(t), \{\mathbf{s}_j(k T_{\mathrm{up}})\}_{j \in \mathcal{N}_i(t)}\}$, where $\mathcal{N}_i(t) = \{j \mid \|\mathbf{p}_i(t) - \mathbf{p}_j(t)\| \le R_{\mathrm{comm}} \}$ is the neighbor set, and $\mathbf{s}_j(k T_{\mathrm{up}})$ represents the stale state received at the last synchronization instant.

Let $C(t) \in [0,1]$ be the instantaneous spatial coverage rate. We maximize the expected cumulative coverage after a topological perturbation at $t_f$ under policy $\pi$, subject to the following constraints for all $i \in \mathcal{N}$:
\begin{subequations}
\label{eq:optimization_problem}
\begin{align}
    \max_{\pi} \quad & \mathbb{E}_{\pi} \bigg[ \sum_{t=t_f}^{T} \gamma^{t-t_f} C(t) \bigg] \label{eq:obj_main} \\
    \text{s.t.} \quad & 1/T_{\mathrm{up}} \le \Omega, \label{eq:signaling_const} \\
    & \mathbb{P}(\text{SINR}_{ij}(t) \ge \Gamma) \ge 1 - \epsilon, \quad \forall j \in \mathcal{N}_i(t) \label{eq:sinr_const} \\
    & \|\mathbf{p}_i(t) - \mathbf{p}_j(t)\| \ge d_{\mathrm{safe}}, \quad \forall j \neq i \label{eq:safety_const} \\
    & \|\mathbf{v}_i(t)\| \le V_{\max}, \quad \|\mathbf{a}_i(t)\| \le A_{\max} \label{eq:kinematic_const}
\end{align}
\end{subequations}
where $\gamma \in [0,1)$ is the discount factor, $T$ is the horizon, $\Omega$ bounds the signaling frequency, $\epsilon$ is the outage threshold, $d_{\mathrm{safe}}$ is the minimum safe distance, and $V_{\max}, A_{\max}$ denote kinematic limits. Sensing and communication ranges are fixed to isolate signaling staleness. Instead of explicitly penalizing \eqref{eq:sinr_const}, connectivity is implicitly maintained because disconnections naturally degrade $C(t)$. Constraints \eqref{eq:safety_const} and \eqref{eq:kinematic_const} act as soft penalties.


\section{Predictive Lightweight Multi-Agent Reinforcement Learning Framework}
\label{sec:proposed_framework}

\begin{figure}[!t]
    \centering    
    \includegraphics[width=.85\linewidth]{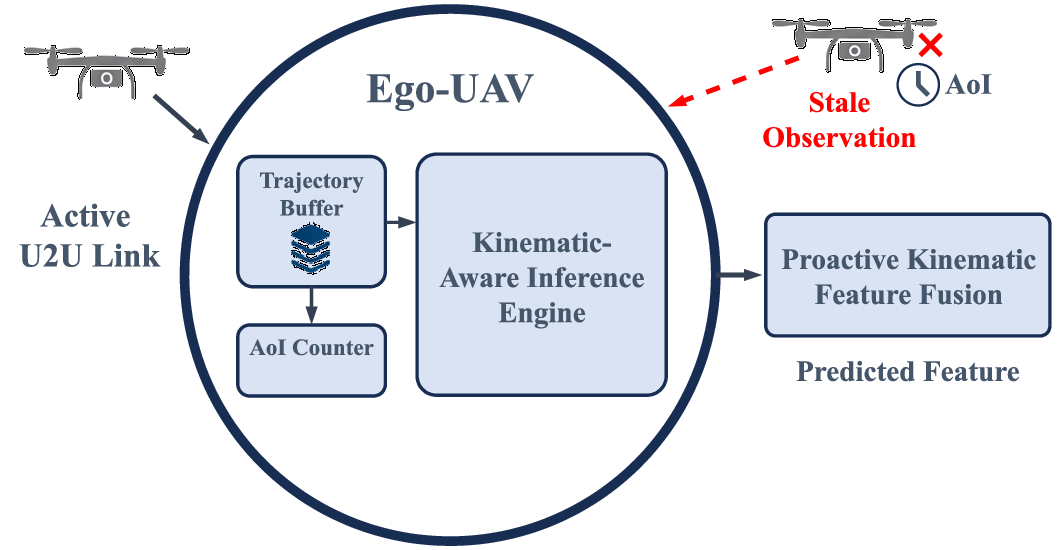}
    \caption{Architecture of the onboard kinematic-aware inference engine. The framework reconstructs physical states $\hat{\mathbf{s}}_{ij}$ from sparse signaling updates to compensate for perception aging, subsequently performing topology-aware feature fusion for decentralized decision-making.}
    \label{fig:inference_engine}
    \vspace{-1em}
\end{figure}

To resolve coordination degradation from stale observations, we propose the PL-MARL framework. Built upon a Digital Twin (DT)-assisted \textit{Centralized Training with Decentralized Execution} (CTDE) paradigm, it explicitly integrates physical kinematic constraints to overcome standard black-box limitations. As illustrated in Fig.~\ref{fig:inference_engine}, the architecture synergistically combines the Kinematic-Aware Inference Engine with proactive kinematic feature fusion, operating as a unified onboard generative engine to enable robust decentralized control with minimal computational overhead.

\subsection{Kinematic-Aware Inference Engine}

During periods of sparse signaling, the delayed perception of neighbor states severely degrades multi-agent coordination. To maintain continuous spatial awareness, node $i$ maintains a local reconstructed trajectory buffer $\overline{\mathcal{H}}_{ij}(t-1) = \{ \hat{\mathbf{s}}_{ij}(t - \tau) \}_{\tau=1}^{L}$, which stores the $L$ most recent states inferred by node $i$ regarding neighbor $j$.
Instead of recurrent sequential unrolling, the $L$ temporal states are concatenated into a single flattened vector and processed by a \textit{Multi-Layer Perceptron} (MLP) in one highly efficient forward pass (hyperparameters detailed in Table \ref{table:parameters}).
The generative predictor $\phi(\cdot)$ operates autoregressively utilizing this flattened buffer and the accumulated P-AoI $\Delta_{ij}(t)$ to estimate the current neighbor position:
\begin{equation}
    \hat{\mathbf{s}}_{ij}(t) = \phi ( \overline{\mathcal{H}}_{ij}(t-1), \Delta_{ij}(t) ; \theta_{\text{inf}} ),
\end{equation}
where $\theta_{\text{inf}}$ denotes the learnable parameters of the inference engine.
Once generated, the newly inferred state $\hat{\mathbf{s}}_{ij}(t)$ is appended to the buffer, replacing the stale component in the decision-making loop. The augmented policy input is thus refined from $\mathcal{O}_i(t)$ to $\hat{\mathcal{O}}_i(t) = \{\mathbf{s}_i(t), \{\hat{\mathbf{s}}_{ij}(t)\}_{j \in \mathcal{N}_i(t)}\}$, ensuring uninterrupted topological reasoning even under extreme signaling scarcity.

To ensure physical plausibility, we formulate trajectory reconstruction as a geometry-constrained estimation problem. Beyond minimizing coordinate reconstruction error, we enforce spatial consistency by constraining the predicted relative displacement $\Delta \hat{\mathbf{p}}_{ij}(t)$ to preserve local topological structure. This yields the objective $\mathcal{L}_{\text{inf}} = \| \mathbf{s}_{j}^{\text{true}}(t) - \hat{\mathbf{s}}_{ij}(t) \|^2 + \lambda \| \text{PE}(\Delta \mathbf{p}_{ij}^{\text{true}}(t)) - \text{PE}(\Delta \hat{\mathbf{p}}_{ij}(t)) \|^2$. The first term ensures trajectory fidelity, while the second term functions as a kinematic-aware regularizer grounding inference in coverage geometry. Here, $\lambda$ is a weighting coefficient, $\text{PE}(\cdot)$ denotes the positional encoding function (detailed in Section III-B), and $\Delta \mathbf{p}_{ij}^{\text{true}}(t)$ represents the true relative displacement at time $t$.

\noindent \textit{Remark:} Ground-truth trajectory information ($\mathbf{s}_{j}^{\text{true}}$) is utilized solely during centralized training within the DT environment as a supervised signal. During decentralized execution, global states are unavailable. UAVs rely exclusively on local observations and pre-trained inference, ensuring compliance with real-world, sparse-signaling operational constraints.

\subsection{Proactive Kinematic Feature Fusion}

Upon reconstructing the local belief states $\hat{\mathbf{s}}_{ij}(t)$, the unified engine performs topology-aware aggregation. Standard feature-based attention mechanisms often neglect spatial proximity, which limits their effectiveness in coverage optimization. To address this, we augment neighbor embeddings with explicit positional information:
\begin{equation}
    \hat{\mathbf{h}}_{ij}(t) = \text{Encoder}(\hat{\mathbf{s}}_{ij}(t)) + \text{PE}(\Delta \mathbf{p}_{ij}(t)),
\end{equation}
where $\text{Encoder}(\cdot)$ extracts the latent semantics and $\Delta \mathbf{p}_{ij}(t) = \hat{\mathbf{s}}_{ij}(t) - \mathbf{s}_i(t)$ is the relative displacement. $\text{PE}(\cdot)$ acts as a learnable projection that maps relative physical displacements into the latent space. This positional encoding ensures that the proactive attention coefficients $e_{ij}(t) = \text{LeakyReLU} (\mathbf{a}^{\mathsf{T}} [ \mathbf{W}\mathbf{h}_i(t) \parallel \mathbf{W}\hat{\mathbf{h}}_{ij}(t) ])$ are kinematically informed by inter-node geometry, where $\mathbf{W}$ is a shared weight matrix, $\mathbf{a}$ is the attention vector, $\parallel$ denotes concatenation, and $\mathbf{h}_i(t)$ is the ego-state embedding. To capture the asymmetric importance, these coefficients are normalized via a softmax operation to obtain $\alpha_{ij}(t) = \exp(e_{ij}(t)) / \sum_{k \in \mathcal{N}_i(t)} \exp(e_{ik}(t))$. Consequently, the aggregated message is computed inline as $\mathbf{m}_i(t) = \sigma \big( \sum_{j \in \mathcal{N}_i(t)} \alpha_{ij}(t) \mathbf{W}\hat{\mathbf{h}}_{ij}(t) \big)$ to fuse both latent semantic features and explicit spatial priors, where $\sigma$ denotes the activation function (e.g., ELU).
This architecture allows the decentralized controller to implicitly prioritize neighbors based on their actual physical impact on network coverage, significantly enhancing the coordination precision in dynamic aerial environments.

\begin{algorithm2e}[!t]
\small
\caption{Kinematic-Aware Decentralized Coordination}
\label{alg:kinematic_coord}
\SetAlgoLined
\KwIn{Time step $t$, Signaling interval $T_{\mathrm{up}}$, Buffer $\overline{\mathcal{H}}_{ij}$, Neighbor index $\mathcal{N}_i(t)$, Ego state $\mathbf{s}_i(t)$}
\KwOut{Coordination action $\mathbf{a}_i(t)$}

\ForEach{neighbor $j \in \mathcal{N}_i(t)$}{
    \eIf{$t \pmod{T_{\mathrm{up}}} == 0$}{
        Receive true state $\mathbf{s}_j(t)$ via signaling update\;
        Update buffer $\overline{\mathcal{H}}_{ij}$ and reset P-AoI $\Delta_{ij}(t) \gets 1$\;
        $\hat{\mathbf{s}}_{ij}(t) \gets \mathbf{s}_j(t)$\;
    }{
        Increment P-AoI $\Delta_{ij}(t) \gets \Delta_{ij}(t) + 1$\;
        \tcp{Reconstruct physical state via inference engine}
        $\hat{\mathbf{s}}_{ij}(t) \gets \phi(\overline{\mathcal{H}}_{ij}(t-1), \Delta_{ij}(t); \theta_{\text{inf}})$\;
    }
    \tcp{Inject positional encoding into feature embedding}
    $\Delta \mathbf{p}_{ij}(t) \gets \hat{\mathbf{s}}_{ij}(t) - \mathbf{s}_i(t)$\;
    $\hat{\mathbf{h}}_{ij}(t) \gets \text{Encoder}(\hat{\mathbf{s}}_{ij}(t)) + \text{PE}(\Delta \mathbf{p}_{ij}(t))$\;
}
Construct spatio-temporal graph $\mathcal{G}(t)$ using $\{\hat{\mathbf{h}}_{ij}(t)\}$\;
Compute attention weights $\alpha_{ij}(t)$ and message $\mathbf{m}_i(t)$\;
Generate action $\mathbf{a}_i(t) \sim \pi(\mathbf{s}_i(t), \mathbf{m}_i(t))$\;
\Return $\mathbf{a}_i(t)$
\end{algorithm2e}

\subsection{Training and Lightweight Efficiency}

Algorithm \ref{alg:kinematic_coord} summarizes the predict-then-coordinate flow. We train the framework via centralized MAPPO within an edge DT. To satisfy Section II constraints and enhance resilience against node or link failures, we employ an adaptive global reward $R(t)$:
\begin{equation}
\label{eq:reward_shaping}
    R(t) = \omega_1 C(t) + \omega_2(f) P(t) - \omega_3(f) D(t) - \omega_4 \Psi_{\text{safe}}(t),
\end{equation}
where $\omega_1$ to $\omega_4$ are scaling weights, $C(t)$ is the coverage rate, $P(t)$ is the potential reward measured by the proximity between the network center and service centroid, and $D(t)$ indicates mobility cost. Crucially, $\Psi_{\text{safe}}(t)$ acts as a repulsive penalty for spatial distance violations below the collision threshold $d_{\mathrm{safe}}$. 
Instead of imposing an explicit penalty for link degradation, we intentionally omit the SINR constraint from the reward formulation to foster emergent collaborative intelligence. Since kinematic-driven feature synthesis relies strictly on neighbors within the communication range $R_{\mathrm{comm}}$, topological disconnections naturally degrade shared observation quality. This compels the model to implicitly discover the correlation between maintaining graph attention connectivity and maximizing the global coverage reward $C(t)$. Consequently, the swarm autonomously self-organizes the optimal trade-off between spatial dispersion for coverage and topological cohesion for feature synthesis, avoiding rigid inductive biases.
We adaptively scale the spatial weights $\omega_2(f)$ and $\omega_3(f)$ based on the topological failure state $f$. Upon detecting a perturbation, the mechanism increases $\omega_2$ to incentivize inward topology contraction, while decreasing $\omega_3$ to facilitate large-scale network reconfiguration.

The computational overhead is primarily dominated by the generative inference and attention mechanisms. Let $M \le N$ represent the average neighborhood size within $R_{\mathrm{comm}}$. To ensure lightweight efficiency, we employ a \textit{Single-Head Attention} (SHA) mechanism. Given a trajectory buffer of length $L$ and a hidden dimension $D$, the local complexity per node is $\mathcal{O}(M(LD + D^2))$, which exhibits linear scaling with respect to $M$ rather than the total swarm size $N$. Although onboard inference introduces marginal processing overhead, it substantially reduces invocations of the energy-intensive radio frequency front-end. Since radio frequency transmission typically dominates the UAV power budget, this computation-for-communication trade-off proves highly efficient for aerial networks under sparse signaling constraints.

\section{Simulation Results and Analysis}
\label{sec:simulation_results}

\subsection{Simulation Setup and Metrics}

We evaluate a $1~\text{km}^2$ 3D coverage scenario with $N=4$ UAVs and 240 Gauss-Markov ground users (max speed $v_u=0.5~\text{m/s}$). Due to space constraints, we focus on the Crowded Urban environment as a rigorous stress test against severe topological perturbations, though preliminary results confirm consistent PL-MARL gains in Suburban and Rural settings. 
To prevent overfitting to specific macro-mobility patterns, training employs a mixed-mobility curriculum (Random Waypoint, Gauss-Markov, and Reference Point Group Mobility).
For realistic onboard processing, we implement a multi-rate architecture. Environment physics and the inference engine operate at a $1~\text{s}$ resolution for continuous tracking, while the decentralized policy updates actions every $\tau=3~\text{s}$ to conserve computation.
Episodes span $200~\text{s}$ ($T=200$ steps). 
Metrics are evaluated over 100 independent episodic runs, with shaded regions in all continuous plots denoting 95\% confidence intervals. Furthermore, performance gains over baselines are strictly validated via the non-parametric Mann-Whitney U test ($p < 0.01$).

To evaluate stringent bandwidth constraints, we vary the signaling interval $T_{\mathrm{up}} \in \{5, 10, \ldots, 50\}$ steps, where $T_{\mathrm{up}}=50$ imposes a severe $50~\text{s}$ silence. Despite large theoretical displacements, swarm connectivity persists as agents proactively bound their mobility to match ground dynamics. A buffer horizon $L=10$ provides the inference engine with high-order motion priors, enabling smooth state fusion during intermittent updates. Furthermore, to foster topological resilience, training is executed for 1000 episodes; after the 500th episode, each episode incorporates a 0.2 probability of a single random node failure at $t=100$. Detailed parameters are listed in Table \ref{table:parameters}.

We compare PL-MARL against three benchmarks: 1) \textbf{TAG-MAPPO (Ablated Baseline)} \cite{lai2026resilienttopologyawarecoordinationdynamic}, which relies solely on stale observations without trajectory inference; 2) \textbf{GRU-based MARL}, which substitutes the inference engine with a standard \textit{Gated Recurrent Unit} (GRU) to evaluate the merit of physical priors; and 3) \textbf{Oracle Bound}, the theoretical coverage ceiling derived via sequential greedy grid-search under perfect global knowledge.

\subsection{Generalization and Resilience to Signaling Scarcity}

\begin{figure}[!t]
    \centering 
    \begin{subfigure}[b]{0.5\columnwidth}
        \includegraphics[width=\linewidth]{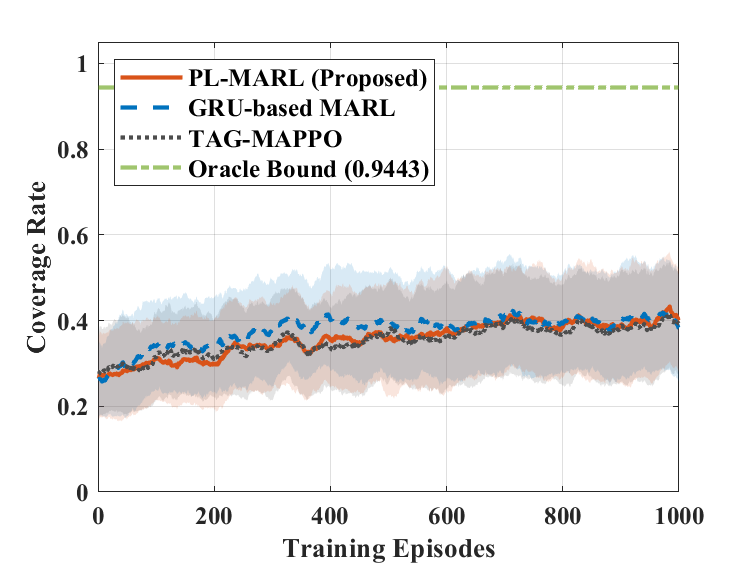}
        \caption{Convergence Analysis}
        \label{fig:convergence_results:a}
    \end{subfigure}\hfill    
    \begin{subfigure}[b]{0.5\columnwidth}
        \centering
        \includegraphics[width=\linewidth]{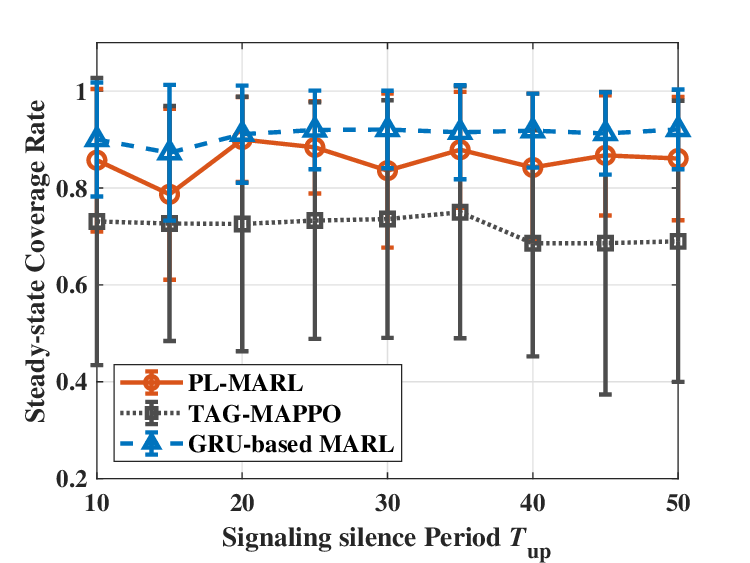}
        \caption{Coverage Rate vs. $T_{\mathrm{up}}$}
        \label{fig:convergence_results:b}
    \end{subfigure}
    \caption{Performance evaluation in the Crowded Urban scenario. \subref{fig:convergence_results:a} Convergence behavior compared against the Oracle Bound. \subref{fig:convergence_results:b} Steady-state coverage under deterministic decision policies across varying $T_{\mathrm{up}}$.}
    \label{fig:convergence_results}
    \vspace{-1.25em}
\end{figure}

We evaluate the Crowded Urban scenario in Fig.~\ref{fig:convergence_results}. Fig.~\ref{fig:convergence_results:a} shows training convergence under stochastic policy exploration. 
Here, continuous action exploration introduces high-entropy trajectory jitter, constraining the average coverage to approximately $0.4$.
Despite severe dynamics, PL-MARL converges stably toward the theoretical Oracle Bound. While the GRU-based MARL baseline outperforms the ablated TAG-MAPPO by leveraging temporal features, it suffers from predictive degradation without explicit physical priors.

Fig.~\ref{fig:convergence_results:b} demonstrates operational resilience under deterministic decision-making with extreme signaling scarcity ($T_{\mathrm{up}}=50$ steps). 
By removing exploratory variance, the swarm successfully exploits learned kinematic priors to stabilize at a higher operational coverage exceeding $0.8$, isolating true robustness from training noise.
Although the GRU baseline achieves a competitive mean coverage, its wide confidence intervals reveal substantial volatility and predictive drift inherent in pure data-driven recurrent architectures. Conversely, PL-MARL tightly bounds predictive uncertainty via kinematic-aware priors, ensuring highly stable spatial reconfiguration under bandwidth-constrained regimes.

\begin{table}[!t]
\centering
\caption{Simulation and Training Parameters}
\label{table:parameters}
\begin{tabular}{@{}ll@{}}
\toprule
\textbf{Parameter} & \textbf{Value} \\ \midrule
Network Area & $1,000~\text{m} \times 1,000~\text{m}$ \\
Number of UAVs ($N$) & $4$ \\
Max. UAV Speed ($V_{\max}$) & $20~\text{m/s}$ \\
Max. UAV Acceleration ($A_{\max}$) & $5~\text{m/s}^2$ \\
Max. Ground User Speed ($v_u$) & $0.5~\text{m/s}$ \\
Comm. / Sensing Radii ($R_{\text{comm}}$ / $R_{\text{s}}$) & $350~\text{m}$ / $100~\text{m}$ \\
Default Signaling Interval ($T_{\text{up}}$) & $15$ steps \\
Action Decision Interval ($\tau$) & $3$ steps \\
Buffer Horizon ($L$) & $10$ steps \\
\midrule
\textbf{Adaptive Reward Weights} & \textbf{Pre- / Post-Perturbation} \\
Coverage Weight ($\omega_1$) & $50.0$ / $50.0$ \\
Potential Weight ($\omega_2$) & $0.5$ / $2.0$ \\
Mobility Cost ($\omega_3$) & $2 \times 10^{-5}$ / $1 \times 10^{-5}$ \\
Repulsive Penalty Weight ($\omega_4$) & $1.0$ / $1.0$ \\
\midrule
Inference Engine MLP Layers / Hidden Dims. & 2 / 64, 32 \\
Fusion Node Embedding Dimension & $32$ \\
Learning Rate (Actor / Critic) & $1 \times 10^{-4}$ / $5\times 10^{-4}$ \\
Total Steps per Episode ($T$) & $200$ \\
Discount Factor / Attention Heads & $0.99$ / $1$ \\
PPO Batch Size / Entropy Schedule & $800$ / $0.05 \rightarrow 0.02$ \\ 
PPO Internal Epochs & $5$ \\
Total Training Episodes & $1,000$ \\
LR Annealing Episodes / Min. Ratio & $800$ / $0.1$ \\
\bottomrule
\end{tabular}
\end{table}

\subsection{Zero-shot Scalability Analysis}

To evaluate zero-shot scalability, the inference engine trained exclusively at $N=4$ is directly deployed up to $N=20$ without retraining. We proportionally scale task clusters alongside the swarm size, maintaining a constant density of 60 users per cluster to isolate coordination efficacy.

\begin{figure}[!t]    
    \centering 
    \begin{subfigure}[b]{0.5\columnwidth}
        \includegraphics[width=\linewidth]{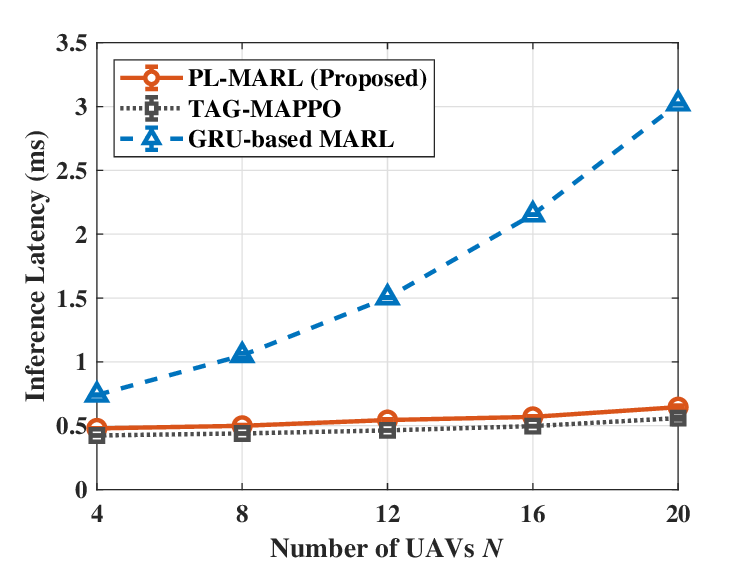}
        \caption{Inference Latency vs. $N$}
        \label{fig:scalability_results:a}
    \end{subfigure}\hfill
    \begin{subfigure}[b]{0.5\columnwidth}
        \centering
        \includegraphics[width=\linewidth]{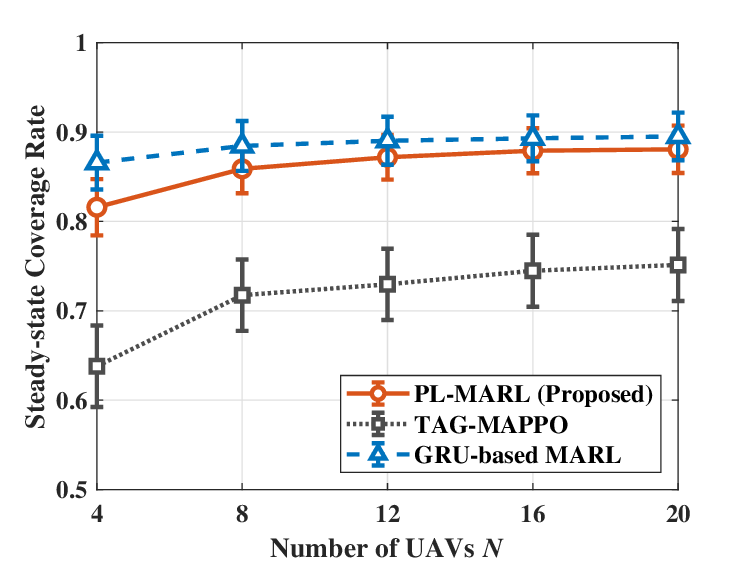}
        \caption{Coverage Rate vs. $N$}
        \label{fig:scalability_results:b}
    \end{subfigure}
    \caption{Zero-shot scalability evaluation of the proposed PL-MARL and baseline models. \subref{fig:scalability_results:a} Inference latency versus the number of UAVs $N$. \subref{fig:scalability_results:b} Steady-state coverage rate versus the number of UAVs $N$.}
    \label{fig:scalability_results}
    \vspace{-.5em}
\end{figure}

Fig.~\ref{fig:scalability_results:a} illustrates the computational demand. The GRU-based MARL exhibits polynomial growth in inference latency, reaching 3.0 ms at $N=20$ due to sequential recurrent unrolling and dense matrix multiplications as the aggregated observation dimension expands. Although a 3.0 ms latency fits within the 3 s control cycle, this non-linear scaling causes severe energy depletion and processor load on resource-constrained micro aerial vehicles. Conversely, PL-MARL maintains a near-constant latency of approximately 0.5 ms by combining a highly parallelizable feedforward architecture with a lightweight SHA mechanism. This decoupling of computational complexity from swarm size makes our framework uniquely suited for battery-limited operations.

Fig.~\ref{fig:scalability_results:b} demonstrates the steady-state coverage rate. While the GRU baseline achieves marginally higher coverage, it incurs substantial computational overhead. PL-MARL provides a superior trade-off, maintaining competitive coverage and minimal latency across all scales. Furthermore, it significantly outperforms TAG-MAPPO, confirming that purely reactive models fail to scale in dynamic environments. This validates PL-MARL as an efficient solution that successfully decouples coordination performance from the scaling bottlenecks inherent in recurrent architectures.

\subsection{Resilience Analysis}

\begin{figure}[!t]    
    \centering 
    \begin{subfigure}[b]{0.5\columnwidth}
        \includegraphics[width=\linewidth]{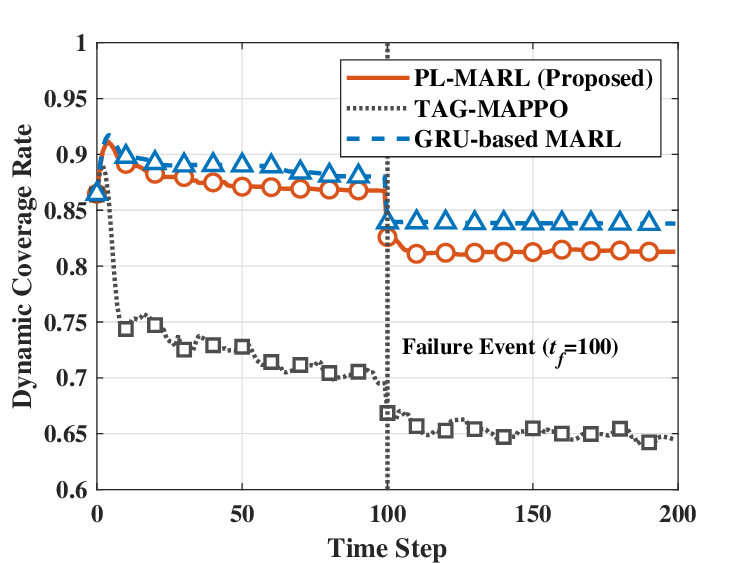}
        \caption{Dynamic Coverage Continuity}
        \label{fig:resilience_results:a}
    \end{subfigure}\hfill
    \begin{subfigure}[b]{0.5\columnwidth}
        \centering
        \includegraphics[width=\linewidth]{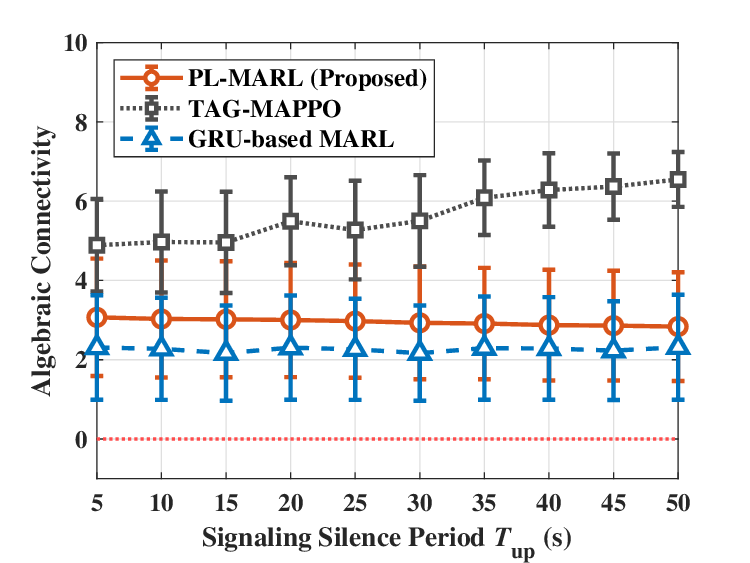}
        \caption{Connectivity ($\lambda_2$) vs. $T_{\mathrm{up}}$}
        \label{fig:resilience_results:b}
    \end{subfigure}
    \caption{Resilience evaluation in the Crowded Urban scenario. (a) Dynamic coverage continuity following node failure at $t=100$. (b) Algebraic connectivity versus signaling silence $T_{\mathrm{up}}$.}
    \label{fig:resilience_results}
    \vspace{-.5em}
\end{figure}

Fig.~\ref{fig:resilience_results} examines autonomous resilience under signaling scarcity. In Fig.~\ref{fig:resilience_results:a}, simulating a random node failure at $t=100$ reveals that PL-MARL maintains graceful coverage degradation. As interpreted directly from the transient curves in Fig.~\ref{fig:resilience_results:a}, the graceful degradation process can be characterized by the initial coverage trough and time-to-stabilization. Specifically, TAG-MAPPO suffers a severe trough (dropping near 0.65) and fails to establish a stable plateau. The GRU baseline delays stabilization by 15 steps. Conversely, PL-MARL actively restricts its trough to approximately 0.81 and rapidly stabilizes to a new steady-state plateau within 10 steps, maximally exploiting the remaining physical capacity. By proactively managing the transient shock and stabilization latency, PL-MARL consistently bridges the gap between static baselines and fragile, high-performing heuristic approaches.

The underlying topological behaviors are exposed in Fig.~\ref{fig:resilience_results:b}, which quantifies integrity via algebraic connectivity ($\lambda_2$). Three distinct regimes emerge: TAG-MAPPO exhibits an overly redundant topology ($\lambda_2 \approx 6.5$), forcing excessive cohesion that restricts spatial exploration. Conversely, GRU-based MARL adopts an extremely sparse, almost disjointed policy ($\lambda_2 \approx 2.2$). While this extreme sparsity inadvertently avoids massive structural reconfigurations during node failure, it operates critically near the threshold of complete network partition, making the system vulnerable to environmental variations. 
PL-MARL maintains an optimal topological balance ($\lambda_2 \approx 3.0$). By deliberately trading unnecessary cohesion for an expanded spatial footprint, it maximizes coverage efficiency without sacrificing network coherence. This confirms that our inference engine successfully decouples structural resilience from signaling frequency, enabling robust, task-driven spatial reconfiguration rather than passive hovering or fragile sparsity.

\section{Conclusion}
\label{sec:conclusion}
This letter proposes the PL-MARL framework for resilient coverage in bandwidth-constrained aerial networks. By integrating a Kinematic-Aware Inference Engine with topology-aware graph attention, we decouple policy execution from communication frequency. Experimental results validate that onboard predictive reasoning mitigates information aging, enabling task-driven spatial reconfiguration over passive, redundant cohesion. Unlike recurrent baselines restricted by polynomial computational scaling and structural fragility, PL-MARL achieves an optimal balance among inference efficiency, energy conservation, and mission reliability. The framework demonstrates a near-constant sub-millisecond latency profile, graceful performance degradation, and robust scalability, proving that predictive physical priors are essential for resource-constrained operations. Consequently, PL-MARL offers a deployment-ready solution for battery-limited and interference-prone aerial missions. By proactively minimizing the signaling footprint, it translates communication constraints into a strategic advantage, ensuring cyber-physical resilience while preserving critical spectrum for user services.

\bibliographystyle{IEEEtran}
\bibliography{Lai_WCL2026-1967_IEEEabrv, Lai_WCL2026-1967_reference}
\ifCLASSOPTIONcaptionsoff  \newpage \fi

\end{document}